\documentclass[12pt]{article}
\usepackage{myart}

\normalbaselines
\input tcilatex

\begin{document}

\author{Yuri A. Rylov}
\title{Contemporary development of Einstein's ideas on space-time and
Brownian motion}
\date{Institute for Problems in Mechanics, Russian Academy of Sciences,\\
101-1, Vernadskii Ave., Moscow, 119526, Russia.\\
e-mail: rylov@ipmnet.ru\\
Web site: {$http://rsfq1.physics.sunysb.edu/\symbol{126}rylov/yrylov.htm$}\\
or mirror Web site: {$http://gasdyn-ipm.ipmnet.ru/\symbol{126}%
rylov/yrylov.htm$}}
\maketitle

\begin{abstract}
Development of the contemporary theory of physical phenomena in the
microcosm is considered to be a result of development of Einstein's ideas on
a possibility of the event space modification and on a possibility of
stochastic (Brownian) motion of free particles. One considers the space-time
modification, based on a new conception of geometry. In the framework of
this conception any geometry is obtained as a result of the proper Euclidean
geometry deformation. In the framework of this conception it is possible
such a space-time geometry, where the free particle motion appears to be
stochastic, although the geometry in itself remains to be deterministic. The
stochasticity intensity depends on the particle mass. It is small for
particle of a large mass, and it is essential for the particle of a small
mass. The space-time geometry may be chosen in such a way, that the
statistical description of random world lines of particles be equivalent to
the quantum description (the Schr\"{o}dinger equation). At such a choice of
the space-time geometry the world function depends on the quantum constant,
and the universal character of the quantum constant is a corollary of the
fact, that all physical phenomena take place in the space-time, whose
properties depend on the value of the quantum constant. The stochastic
motion of free particles may be considered to be a kind of the relativistic
Brownian motion, whose properties are conditioned by the properties of the
space-time geometry. At such a description the quantum principles are not
used. They can be obtained as a corollaries of the statistical description
of nonrelativistic stochastic particles. The nonrelativistic quantum
principles may not be used for description of relativistic phenomena. They
are to be modified. This modification should be obtained from the
statistical description of the relativistic stochastic particles.
\end{abstract}

\section{Introduction}

Papers \cite{E1905a,E1905b} by A. Einstein on modification of the event
space model and the Brownian particle motion concerned the most actual
physical problems of the beginning of the 20th century. Elimination of the
absolute simultaneity, suggested by A. Einstein, generated a modification of
the event space. Isaac Newton considered the event space as the direct
product of time and of the three-dimensional space. A. Einstein suggested to
replace the Newtonian event space by united event space (space-time) with
the common geometry. In his paper \cite{E1905a} he formulated first the idea
on possibility and necessity of the event space modification. This idea,
known as a revision of the space-time conception, is connected closely with
the general direction of the theoretical physics development -- the
increasing geometrization of physics and increasing role of the space-time
structure in the explanation of physical phenomena. The first modification
of the event space structure, produced by A. Einstein and the construction
of the space-time geometry, carried out by G. Minkowski made a start for
further modifications of the space-time. The second modification of the
space-time was produced by A. Einstein in the beginning of the 20th century.
This modification admitted the space-time curvature, which was generated by
the matter placed in the space-time. The second modification, known as the
general relativity, is connected with the supposition that the space-time
geometry may be described by the Riemannian geometry.

In the beginning of the 20th century one discovered that the free particles
of small mass move stochastically. The motion of free particles depends only
on the space-time properties, and one needs the third modification of the
space-time geometry. One needed such a space-time geometry, where the free
particle motion be primordially stochastic and the particle mass be
geometrized. Such geometry is impossible in the framework of the Riemannian
geometry, which was used for the space-time geometry in the beginning of the
20th century.

The third modification of the space-time, has been produced only in the end
of the 20th century. It was generated by appearance of the non-Riemannian
geometry (T-geometry), possessing the earlier unknown property, that in the
space-time with such a geometry the free particle motion is primordially
stochastic (random). This stochasticity has no other reason, except for the
space-time properties. Besides, the space-time with T-geometry contains an
\textquotedblright elementary length\textquotedblright , determined by the
quantum constant, and the intensity of the stochasticity depends on the
particle mass (on ratio of the elementary length to the particle geometrical
mass). The use of T-geometry as a space-time geometry admits one to refuse
the idea about enigmatic quantum nature of the matter and to explain the
quantum phenomena as natural manifestation of the space-time properties.

The Einstein's idea on possibility of the random particle motion without a
construction of a model of the phenomena, generating incident, was advanced
in the paper \cite{E1905b}. The Brownian motion was considered to be
dissipative and nonrelativistic. At the further development of the
stochastic particle motion investigation one discovered that the individual
quantum particle is stochastic (in a sense Brownian). However, the
\textquotedblright Brownian motion\textquotedblright\ of the quantum
particles appeared to be \textit{nondissipative and relativistic}. The
random component of the quantum particle motion is always relativistic, even
in the case, when the regular component of the particle motion is
nonrelativistic. This circumstance is very important, because the
statistical description of the random relativistic motion of a particle
distinguishes conceptually from the nonrelativistic random motion.

At this point we face with the importance of the true understanding of the
relativity theory and application of this understanding to the construction
of the statistical description. It is a common practice to consider, that
for taking into account the relativity theory at a description of a physical
phenomenon it is sufficient that the dynamical equations describing this
phenomenon can be written in the relativistically covariant form. Indeed,
this condition is necessary for the true relativistic description, but it is
not sufficient. For instance, describing the free particle motion, it is
important not only to write dynamical equations correctly. It is necessary
more to point out correctly what is the particle state. The classics of the
relativity theory knew this circumstance \cite{F55}. In the nonrelativistic
theory the state of several particles is described by their positions and
momenta, i.e. by some point in the phase space. The values of positions and
momenta are taken at the same time moment, which is common for all
particles. It means that the physical object is a particle, i.e. a point in
the three-dimensional space of positions.

In the relativistic theory one may not describe the state of several
particles by a point in the phase space, i.e. by positions and momenta taken
at the same time moment, as far as there is no absolute simultaneity.
Positions and momenta of several particles taken at the same time correspond
to different sets of events in different coordinate systems. In the case of
deterministic particles the positions and momenta, taken at the same time in
one inertial coordinate system, can be recalculated to the moving inertial
coordinate system, if the particle are deterministic and their dynamic
equations are known. However, one cannot do this in the case of stochastic
particles, for which there are no dynamic equations. It is a reason, why the
state of a relativistic particle is described by the world line. The world
line (but not a point in the phase space) is the physical object, liable to
the statistical description.

\section{Statistical description of relativistic particles}

Any statistical description is a description of \textit{physical objects}.
In the nonrelativistic case the physical objects are points of the
three-dimensional space. In the relativistic case the physical objects are
lengthy objects: world lines. Statistical description of nonrelativistic
particles distinguishes from that of relativistic particles in the relation,
that the state density $\rho $ at the nonrelativistic description is defined
by the relation 
\begin{equation}
dN=\rho dV  \label{a1.1}
\end{equation}
whereas in the relativistic case the state density $j^{k}$ is described by
the relation 
\begin{equation}
dN=j^{k}dS_{k}  \label{a1.2}
\end{equation}
where $dN$ is the flux of world lines through the infinitesimal area $dS_{k}$%
. It follows from the relations (\ref{a1.1}) 
\"{}
(\ref{a1.2}) that in the nonrelativistic case one can introduce the concept
of the probability density of the state on the basis of the nonnegative
quantity $\rho $, whereas in the relativistic case it is impossible, because
one cannot construct the probability density on the basis of the 4-vector $%
j^{k}$.

Statistical description is a description of the statistical ensemble, i.e.
the dynamic system consisting of many identical independent systems. These
systems may be dynamical or stochastic. However, the statistical ensemble is
a dynamic system in any case. It means, that there are dynamic equations,
which describe the evolution of the statistical ensemble state.
Investigation of the statistical ensemble as a dynamic system admits one to
investigate the mean characteristics of the stochastic systems, constituting
this ensemble. Besides, in the nonrelativistic case the statistical ensemble
is a tool for calculation of different mean quantities and distributions,
because in this case the ensemble state may be interpreted as the
probability density of the fact, that the system state is placed at some
given point of the phase space.

The statistical ensemble is used usually in the statistical physics, where
the statistical description of the deterministic nonrelativistic systems is
produced. The principal property of the statistical ensemble (\textit{to be
a dynamic system}) is perceived as some triviality, and the statistical
ensemble is considered usually as a tool for calculation of mean values of
different functions of the state. When one tries to apply this conception of
the statistical ensemble to description of relativistic stochastic
particles, it is quite natural that one fails, because the probabilistic
conception of the statistical ensemble (statistical ensemble as a tool for
calculation of mean values) cannot be applied here. The problem of
construction of a dynamic system (statistical ensemble) from stochastic
systems is not stated simply.

We display in the example of free nonrelativistic particles, how the
statistical ensemble is constructed. The action $\mathcal{A}_{\mathcal{S}_{%
\mathrm{d}}}$ for the free deterministic particle $\mathcal{S}_{\mathrm{d}}$
has the form 
\begin{equation}
\mathcal{A}_{\mathcal{S}_{\mathrm{d}}}\left[ \mathbf{x}\right] =\int \frac{m%
}{2}\left( \frac{d\mathbf{x}}{dt}\right) ^{2}dt  \label{c3.7a}
\end{equation}%
where $\mathbf{x}=\mathbf{x}\left( t\right) $. For the statistical ensemble $%
\mathcal{A}_{\mathcal{E}\left[ \mathcal{S}_{\mathrm{d}}\right] }$ of free
deterministic particles we obtain the action 
\begin{equation}
\mathcal{A}_{\mathcal{E}\left[ \mathcal{S}_{\mathrm{d}}\right] }\left[ 
\mathbf{x}\right] =\int \frac{m}{2}\left( \frac{d\mathbf{x}}{dt}\right)
^{2}dtd\mathbf{\xi }  \label{c3.7}
\end{equation}%
where $\mathbf{x}=\mathbf{x}\left( t,\mathbf{\xi }\right) $ is a 3-vector
function of independent variables $t,\mathbf{\xi =}\left\{ \xi _{1,}\xi
_{2},\xi _{3}\right\} $. The variables (Lagrangian coordinates) $\mathbf{\xi 
}$ label particles $\mathcal{S}_{\mathrm{d}}$ of the statistical ensemble $%
\mathcal{E}\left[ \mathcal{S}_{\mathrm{d}}\right] $. The statistical
ensemble $\mathcal{E}\left[ \mathcal{S}_{\mathrm{d}}\right] $ is a dynamic
system of hydrodynamic type.

The statistical ensemble $\mathcal{E}\left[ \mathcal{S}_{\mathrm{st}}\right] 
$ of free \textit{stochastic} particles $\mathcal{S}_{\mathrm{st}}$ is a
dynamical system, described by the action 
\begin{equation}
\mathcal{A}_{\mathcal{E}\left[ \mathcal{S}_{\mathrm{st}}\right] }\left[ 
\mathbf{x,u}_{\mathrm{df}}\right] =\int \left\{ \frac{m}{2}\left( \frac{d%
\mathbf{x}}{dt}\right) ^{2}+\frac{m}{2}\mathbf{u}_{\mathrm{df}}^{2}-\frac{%
\hbar }{2}\mathbf{\nabla u}_{\mathrm{df}}\right\} dtd\mathbf{\xi }
\label{c4.1}
\end{equation}%
where $\mathbf{u}_{\mathrm{df}}=\mathbf{u}_{\mathrm{df}}\left( t,\mathbf{x}%
\right) $ is a diffusion velocity, describing the mean value of the
stochastic component of velocity, whereas $\frac{d\mathbf{x}}{dt}\left( t,%
\mathbf{\xi }\right) $ describes the regular component of the particle
velocity, and $\mathbf{x}=\mathbf{x}\left( t,\mathbf{\xi }\right) $ is the
3-vector function of independent variables $t,\mathbf{\xi =}\left\{ \xi
_{1,}\xi _{2},\xi _{3}\right\} $. The variables $\mathbf{\xi }$ label
particles $\mathcal{S}_{\mathrm{st}}$, substituting the statistical
ensemble. The operator 
\[
\mathbf{\nabla =}\left\{ \frac{\partial }{\partial x^{1}},\frac{\partial }{%
\partial x^{2}},\frac{\partial }{\partial x^{2}}\right\} 
\]
is defined in the coordinate space $\mathbf{x}$.

The action for the single stochastic particle is obtained from the action (%
\ref{c4.1}) by omitting integration over $\mathbf{\xi }$. However, the
obtained action 
\begin{equation}
\mathcal{A}_{\mathcal{S}_{\mathrm{st}}}\left[ \mathbf{x,u}_{\mathrm{df}}%
\right] =\int \left\{ \frac{m}{2}\left( \frac{d\mathbf{x}}{dt}\right) ^{2}+%
\frac{m}{2}\mathbf{u}_{\mathrm{df}}^{2}-\frac{\hbar }{2}\mathbf{\nabla u}_{%
\mathrm{df}}\right\} dt  \label{c4.1a}
\end{equation}
has only symbolic sense, as far as the operator $\mathbf{\nabla }$ is
defined in some vicinity of the point $\mathbf{x}$, but not at the point $%
\mathbf{x}$ itself. It means, that the action (\ref{c4.1a}) does not
determine dynamic equations for the particle $\mathcal{S}_{\mathrm{st}}$,
and the particle appears to be stochastic, although dynamic equations for
the statistical ensemble of such particles exist. They are determined by the
action (\ref{c4.1}). Thus, the particles described by the action (\ref{c4.1}%
) are stochastic, because there are no dynamic equations for a single
particle. In the case, when the quantum constant $\hbar =0$, the actions (%
\ref{c4.1a}) and (\ref{c3.7a}) coincide, because in this case it follows
from (\ref{c4.1a}), that $\mathbf{u}_{\mathrm{df}}=0$.

Variation of action (\ref{c4.1}) with respect to variable $\mathbf{u}_{%
\mathrm{df}}$ leads to the equation 
\begin{equation}
\mathbf{u}_{\mathrm{df}}=-\frac{\hbar }{2m}\mathbf{\nabla }\ln \rho ,
\label{c4.5}
\end{equation}
where the particle density $\rho$ is defined by the relation 
\begin{equation}
\rho =\left[ \frac{\partial \left( x^{1},x^{2},x^{3}\right) }{\partial
\left( \xi _{1},\xi _{2},\xi _{3}\right) }\right] ^{-1}=\frac{\partial
\left( \xi _{1},\xi _{2},\xi _{3}\right) }{\partial \left(
x^{1},x^{2},x^{3}\right) }  \label{c4.4}
\end{equation}
The relation (\ref{c4.5}) is the expression for the mean diffusion velocity
in the Brownian motion theory.

Eliminating $\mathbf{u}_{\mathrm{df}}$ from the dynamic equation for $%
\mathbf{x}$, we obtain dynamic equations of the hydrodynamic type. 
\begin{equation}
m\frac{d^{2}\mathbf{x}}{dt^{2}}=-\mathbf{\nabla }U\left( \rho ,\mathbf{%
\nabla }\rho \right) ,\qquad U\left( \rho ,\mathbf{\nabla }\rho \right) =%
\frac{\hbar ^{2}}{8m}\left( \frac{\left( \mathbf{\nabla }\rho \right) ^{2}}{%
\rho ^{2}}-2\frac{\mathbf{\nabla }^{2}\rho }{\rho }\right)  \label{c4.7}
\end{equation}
By means of the proper change of variables these equations can be reduced to
the Schr\"odinger equation \cite{R99}.

However, there is a serious mathematical problem here. The fact is that the
hydrodynamic equations are to be integrated, in order they can be described
in terms of the wave function. The Schr\"{o}dinger equation consists of two
real equations. To obtain from them four hydrodynamic equations, one needs
to take gradient from one of real components of the Schr\"{o}dinger equation
and to introduce proper designations. The inverse transition from
hydrodynamic equations to their representation in terms of the wave function
needs an integration. Three arbitrary functions appear as a result of this
integration, and two-component wave function is constructed from these
arbitrary functions \cite{R99}. The fact, that the Schr\"{o}dinger equation
can be written in the hydrodynamic form, is well known \cite{M26,B52}.
However, the inverse transition from the hydrodynamic equations to the wave
function was not known until the end of the 20th century \cite{R99}, and the
necessity of integration of hydrodynamic equations was a reason of this fact.

Derivation of the Schr\"{o}dinger equation as a partial case of dynamic
equations, describing the statistical ensemble of random particles (\ref%
{c4.1}), shows that the wave function is simply a method of description of
hydrodynamic equations, but not a specific quantum object, whose properties
are determined by the quantum principles. At such an interpretation of the
wave function the quantum principles appear to be superfluous, because they
are necessary only for explanation, what is the wave function and how it is
connected with the particle properties. All remaining information is
contained in the dynamic equations. It appears that the quantum particle is
kind of stochastic particle, and all its exhibitions can be interpreted
easily in terms of the statistical ensemble of stochastic particles (\ref%
{c4.1}).

The idea of that, the quantum particle is simply a stochastic particle, is
quite natural. It was known many years ago \cite{M49}. However, the
mathematical form of this idea could not be realized for a long time because
of the two problems considered above (incorrect conception on the
statistical ensemble of relativistic particles and necessity of integration
of the hydrodynamic equations).

\section{Necessity of the next modification of the space-time model}

Thus, the quantum mechanics can be founded as a mechanics of stochastic
particles. However, it is not known, why the motion of free particles is
stochastic and from where the quantum constant appears. There are two
variants of answer to these questions.

1. The stochasticity of the free particle motion is explained by the
space-time properties, and the quantum constant is a parameter, describing
the space-time properties.

2. The stochasticity of the free particle motion is explained by the special
quantum nature of particles. The motion of such a particle distinguishes
from the motion of usual classical particle. There is a series of rules
(quantum principles), determining the quantum particle motion. The universal
character of the quantum constant is explained by the universality of the
quantum nature of all particles and other physical objects. As to event
space, it remains to be the same as at Isaac Newton.

It is quite clear that the first version of explanation is simpler and more
logical, as far as it supposes \textit{only a change of the space-time
geometry}. The rest, including the principles of classical physics, remains
to be unchanged. The main problem of the first version was an absence of the
space-time geometry with such properties. In general, one could not imagine
that such a space-time geometry can exist. As a result in the beginning of
the 20th century one chose the second version. After a large work the
necessary set of additional hypotheses (quantum principles) had been
invented. One succeeded to explain all nonrelativistic quantum phenomena.
However, an attempt of the quantum theory expansion to the relativistic
phenomena lead to the problem, which is formulated as \textit{join of
nonrelativistic quantum principles with the principles of the relativity
theory}.

The choice of the proper space-time geometry appeared to be possible only
after a change of the geometry conception, which determines what
(generalized) geometries are possible, in general. The new conception of
geometry appeared to be very simple. Any generalized geometry is a
modification of the proper Euclidean geometry. Constructing a generalized
geometry, one repeated conventionally the reasonings of Euclid. At these
reasonings one uses another axioms and deform the space in addition. In the
new conception one suggests to obtain generalized geometries by means of a
simple deformation of the space.

Any geometry is a totality of all geometrical objects and relations between
them. The algorithms of the geometrical objects construction and those of
relation between them has been developed in the proper Euclidean geometry.
Besides, the principles of these algorithms construction have been developed
and compatibility of all axioms, lying in the foundation of these principles
has been proved \cite{H30}. All this admits one to consider the proper
Euclidean geometry as already constructed geometry. Conventionally one uses
a modification of the Euclidean algorithms for a construction of the
generalized geometry. In the new conception of the geometry the modification
of the Euclidean algorithms is simplified essentially. In the new geometry
conception only operand of the Euclidean algorithm is different in different
generalized geometries, but the algorithm in itself remains to be the same
for all generalized geometries.

In order such a representation of Euclidean algorithms to be possible, it is
necessary to use as the operand such a quantity, which determine the
generalized geometry completely. In particular, such an operand is to
determine the proper Euclidean geometry. Such a quantity is the world
function $\sigma =\frac{1}{2}\rho ^{2}$, where the quantity $\rho $ is the
distance between two points of the space \cite{S60}. The circumstance, that
the distance (or the world function) is an important geometrical quantity
was known long ago. But the fact that the world function is the \textit{%
quantity, which determines the geometry completely}, became to be known only
in the end of the 20th century \cite{R90,R2002}. It was proved, that the
proper Euclidean geometry is determined completely by its world function $%
\sigma _{\mathrm{E}}$, if it satisfies the set of conditions, written in
terms of $\sigma _{\mathrm{E}}$. It means, that the algorithm of any
proposition $\mathcal{S}$ of the proper Euclidean geometry $\mathcal{G}_{%
\mathrm{E}}$ can be represented in the form $S\left( \sigma _{\mathrm{E}%
}\right) $, where $S$ is the Euclidean algorithm of construction of the
proposition $\mathcal{S}$. Then the analogous proposition $\mathcal{S}$ of
the generalized geometry $\mathcal{G}$ is described by the same algorithm $%
S\left( \sigma \right) $, but with other operand $\sigma $, which is the
world function of the geometry $\mathcal{G}$.

As far as the construction algorithms of all propositions of the proper
Euclidean geometry are supposed to be known, the construction algorithm of
all propositions of a generalized geometry appears to be very simple. Using
this algorithm, one can construct any $\sigma $-immanent geometry, i.e. the
geometry described completely by its world function $\sigma $. The class of
such geometries is very wide. It is much more powerful, than the class of
the Riemannian geometries, which are used usually for description of the
space-time properties. We shall refer to the $\sigma $-immanent geometries
as the physical geometries, because such geometries are very convenient for
description of the space-time, whose main characteristic is interval between
two events (points in the space-time). Any change of the space-time interval
is accompanied by a change of the world function, and it means some
deformation of the space-time.

The $\sigma $-immanent geometry is called also the tubular geometry (shortly
T-geometry), because in such a geometry a hallow tube plays the role of the
straight line. A tube as a generalization of the one-dimensional straight is
the general case of T-geometry. The proper Euclidean geometry, where the
tube degenerates into one-dimensional straight line, is a very special
(degenerate) case of T-geometry. If one constructs the generalized geometry
on the basis of the deformation, which does not violate one-dimensionality
of the Euclidean straight (the Riemannian geometry is constructed like
this), one may construct only degenerate geometry with one-dimensional lines
instead of tubes. The stochastic motion of free particles is characteristic
for the nondegenerate space-time T-geometry. It is connected with the fact,
that the 4-momentum of a free particle is transferred along the world line
in parallel, and simultaneously it is tangent to the world line, i.e. the
momentum 4-vector determines the world line direction. If there is only one
direction parallel to the direction of the momentum vector at the point $P$,
the direction of the momentum vector at the neighbor point $P^{\prime }$ is
determined single-valuedly, and the world line of the particle appears to be
deterministic. If there are many directions parallel to the momentum vector
at the point $P$, the direction of the momentum vector at the neighbor point 
$P^{\prime }$ appears to be multivarious, and the world line appears to be
random, because the direction of the tangent vector appears to be random.
The straight is defined in the proper Euclidean geometry (as in any
T-geometry) as a set of such points $R$, that the vector $\mathbf{P}_{0}%
\mathbf{R}$ is parallel to the vector $\mathbf{P}_{0}\mathbf{P}_{1}$, which
determines the direction of the straight, passing through the points $P_{0}$%
, ${P}_{1}$, given this straight.

Transformation of the one-dimensional Euclidean straight into a tube is
connected with the circumstance, that in the nondegenerate T-geometry there
are directions (vectors) which are parallel to the given vector, but
nonparallel between themselves. In other words, the parallelism property is
intransitive, in general. In the framework of the Riemannian geometry there
is only one homogeneous isotropic space-time geometry. It is the Minkowski
geometry. The set of all homogeneous isotropic space-time geometries is
described by the set of functions of one argument. Such T-geometries are
described by the world function of the form 
\begin{equation}
\sigma =\sigma _{\mathrm{M}}+D\left( \sigma _{\mathrm{M}}\right)
\label{a3.1}
\end{equation}%
where $\sigma _{\mathrm{M}}$ is the world function of the Minkowski
geometry, and $D\left( \sigma _{\mathrm{M}}\right) $ is the distortion
function, which vanishes for the Minkowski geometry. The free particle
motion in such a space-time geometry is stochastic, in general. It is
deterministic only in the case, when $D\left( \sigma _{\mathrm{M}}\right) =0$
and the space-time T-geometry turns into the Minkowski geometry.

\section{Contemporary state of the theory of physical phenomena in microcosm}

The scheme of a fundamental physical theory is shown in the figure. This
theory is a logical structure. The fundamental principles of the theory are
shown below. The experimental data, which are to be explained by the theory
are placed on high. Between them there are a set of logical corollaries of
the fundamental principles. It is possible such a situation, when for some
conditions one can obtain a list of logical corollaries, placed near the
experimental data. It is possible such a situation, when some circle of
experimental data and physical phenomena may be explained and calculated on
the basis of this list of corollaries without a reference to the fundamental
principles. In this case the list of corollaries of the fundamental
principles may be considered as an independent physical theory. Such a
theory will be referred to as the truncated theory, because it explains not
all phenomena, but only a restricted circle of these phenomena (for
instance, only nonrelativistic phenomena). Examples of truncated physical
theories are known in the history of physics. For instance, the
thermodynamics is such a truncated theory, which is valid only for the
quasistatic thermal phenomena. The thermodynamics is an axiomatic theory. It
cannot be applied to nonstatic thermal phenomena. In this case one should
use the kinetic theory, which is a more fundamental theory, as far as it is
applied both to quasistatic and nonstatic thermal phenomena. Besides, the
thermodynamics can be derived from the kinetic theory as a partial case.

The truncated theory has a set of properties, which provide its wide
application.

Firstly, the truncated theory is simpler, than the fundamental one, because
a part of logical reasonings and mathematical calculations of the
fundamental theory are used in the truncated theory in the prepared form.
Besides, the truncated theory is located near experimental data, and one
does not need long logical reasonings for its application.

Secondly, the truncated theory is a list of prescriptions, and it is not a
logical structure in such extent, as the fundamental theory is a logical
structure. The truncated theory is axiomatic, it contains more axioms, than
the fundamental theory, as far as logical corollaries of the fundamental
theory appear in the truncated theory as fundamental principles (axioms).

Thirdly, being simpler, the truncated theory appears before the fundamental
theory, whose corollary it is. It is a reason of conflicts between the
advocates of the fundamental theory and advocates of the truncated theory,
because the last consider the truncated theory to be the fundamental one.
Such a situation took place, for instance, at becoming of the statistical
physics, when advocates of the axiomatic thermodynamics oppugn against Gibbs
and Boltzmann. Such a situation took place at becoming of the doctrine of
Copernicus-Galileo-Newton, when advocates of the Ptolemaic doctrine oppugn
against the doctrine of Copernicus-Galileo-Newton. They referred that there
was no necessity to introduce the Copernican doctrine, as far as the
Ptolemaic doctrine is simple and customary. Only discovery of the Newtonian
gravitation law and consideration of celestial phenomena, which cannot be
described in the framework of the Ptolemaic doctrine, terminated the contest
of the two doctrines.

The main defect of the truncated theory is an impossibility of its expansion
over wider circle of physical phenomena. For instance, let the truncated
theory explains nonrelativistic physical phenomena. It means, that the basic
propositions of the truncated theory are obtained as corollaries of the
fundamental principles and nonrelativistic character of the considered
phenomena. To expand the truncated theory on relativistic phenomena, one
needs to separate, what in the principles of the truncated theory is a
corollary of fundamental principles and what is a corollary of
nonrelativistic character of the considered phenomena. A successful
separation of the two factors means essentially a perception of the theory
truncation and construction of the fundamental theory. If the fundamental
theory has been constructed, the expansion of the theory on the relativistic
phenomena is obtained by an application of the fundamental principles to the
relativistic phenomena. The obtained theory will describe the relativistic
phenomena correctly. It may distinguish essentially from the truncated
theory, which is applicable for description of only nonrelativistic
phenomena.

The conventional nonrelativistic quantum theory is a truncated theory, which
is applicable for description of nonrelativistic phenomena only. This
statement is called in question usually by researchers working in the field
of the quantum theory. The problem of the relativistic quantum theory is
formulated usually as a problem of unification of the nonrelativistic
quantum principles with the principles of the relativity theory.
Conventionally the nonrelativistic quantum theory is considered to be a
fundamental theory. The relativistic quantum theory is tried to be
constructed without puzzling out, what in the nonrelativistic quantum theory
is conditioned by principles and what is conditioned by its nonrelativistic
character. It is suggested that the linearity is the principle property of
the quantum theory, and it is tried to be saved. However, the analysis shows
that the linearity of the quantum theory is some artificial circumstance 
\cite{R2005}, which simplifies essentially the description of quantum
phenomena, but it does not express the essence of these phenomena. The
conventional approach to construction of the relativistic quantum theory is
shown by the dashed line in the scheme. Following this line, the
construction of the true relativistic quantum theory is as difficult, as
discovery of the Newtonian gravitation law on the basis of the Ptolemaic
conception. Besides, even we succeed to construct such a theory, it will be
very difficult to choose the valid version of the theory, because it has no
logical foundation. In other words, the conventional approach of
construction of the relativistic quantum theory (invention of new hypotheses
and fitting) seems to lead to blind alley, although one cannot eliminate the
case that it appears to be successful.

Alternative way of construction of the relativistic theory of physical
phenomena in the microcosm is shown by the solid line in scheme. It supposes
derivation of fundamental principles and their subsequent application to the
relativistic physical phenomena. Elimination of the nonrelativistic quantum
principles is characteristic for this approach. This elimination is
accompanied by the elimination of the problem of the unification of the
nonrelativistic quantum principles with the relativity principles.
Simultaneously one develops dynamical methods of the quantum system
investigation, when the quantum dynamic system is investigated simply as a
dynamic system. These methods are free of application of quantum principles.
They are used for investigation of both relativistic and nonrelativistic
quantum systems. A use of logical construction is characteristic for this
approach. Invention of new hypotheses and fitting are not used.

Application of dynamical methods to investigation of the Klein-Gordon
particle admits one to discover a special quantum field responsible for the
particle production \cite{R2003}. It is especially important, because in the
classical physics such fields are not known. Application of dynamical
methods to investigation of the Dirac particle admits one to establish, that
the Dirac particle has an internal structure \cite{R2004a}, which is
described nonrelativistically \cite{R2004b}. Application of dynamic methods
admits one to establish the interpretation of quantum phenomena, founded on
the concept of the classical stochastic particle. As a result the
multiplicity of interpretation of the wave function and other quantum
phenomena has been eliminated. In particular, it was shown, that the
Copenhagen interpretation, where the wave function describes an individual
particle, is incompatible with the quantum mechanics formalism \cite%
{R2004c,R2005}. These results cannot be obtained by the conventional
methods, whose capacities are restricted by a use of the quantum principles.
Thus, at construction of relativistic theory of physical phenomena in
microcosm some optimism appears. It is generated by the derivation of
fundamental principles and by application of the dynamic methods of
investigation.

\end{document}